\documentclass[twocolumn,aps,amsmath,amssymb,floatfix]{revtex4}
\usepackage{graphicx}
\usepackage{dcolumn}
\usepackage{bm}
\usepackage{setspace}
\usepackage{color}

\begin{document}

\title{Biophysical mechanism for Ras-nanocluster formation and
signaling in plasma membrane}

\author{Thomas Gurry$^{1,2}$, Ozan Kahramano{\u g}ullar{\i}$^{1,3}$, Robert G. Endres$^{1,4,*}$}
\affiliation{
$^{1}$Centre for Integrated Systems Biology at Imperial College,
Imperial College London, London SW7 2AZ, United Kingdom,\\
$^{2}$Department of Applied Mathematics and Theoretical Physics, University of Cambridge, 
Cambridge CB3 0WA, United Kingdom,\\
$^{3}$Department of Computing, Imperial College London, London SW7 2AZ, United Kingdom,\\
$^{4}$Division of Molecular Biosciences, Imperial College London, London SW7 2AZ\\
$^{*}$To whom correspondence should be addressed. E-mail: r.endres@imperial.ac.uk\\
}

\date{\today}

\begin{abstract}
Ras GTPases are lipid-anchored G proteins which play a fundamental role in cell 
signaling processes.  Electron micrographs of immunogold-labeled Ras have shown 
that membrane-bound Ras molecules segregate into nanocluster domains.  Several models 
have been developed in attempts to obtain quantitative descriptions of nanocluster 
formation, but all have relied on assumptions such as a constant, expression-level 
independent ratio of Ras in clusters to Ras monomers (cluster/monomer ratio).
However, this assumption is inconsistent with the law of mass action.  Here, 
we present a biophysical model of Ras clustering based on short-range attraction 
and long-range repulsion between Ras molecules in the membrane. To test this model, 
we performed Monte Carlo simulations and compared statistical clustering properties 
with experimental data.  We find that we can recover the experimentally-observed 
clustering across a range of Ras expression levels, without assuming a constant cluster/monomer 
ratio or the existence of lipid rafts. In addition, our model makes predictions 
about the signaling properties of Ras nanoclusters in support of the idea that 
Ras nanoclusters act as an analog-digital-analog converter for high fidelity signaling.
\end{abstract}

\maketitle

\section{Introduction}
Plasma membrane heterogeneity is a key concept in molecular cell biology due to its role 
in protein sorting and specificity of signaling \cite{Pouyssegur02,Marguet06,Morone06}. 
Although the diversity of the membrane's lipid components is partly responsible 
for this heterogeneity \cite{Sengupta07}, the role played by membrane proteins is less well understood.  
Members of the Ras protein superfamily \cite{Bar00,Symons01} have been observed to 
form dynamic, non-overlapping domains called nanoclusters in the inner leaflet of the 
plasma membrane \cite{Roy99,Prior01,Prior03,Eisenberg08}.  
While the lateral segregation of Ras may provide evidence towards the 
existence of small, dynamic rafts \cite{Plowman05}, the definition and even existence 
of rafts remains disputed \cite{Munro03}.
In addition to its connection to the lipid-raft concept, Ras has attracted immense
interest due to its fundamental role in a multitude of cellular processes, including
cell proliferation, survival, and motility. Most importantly, 
Ras genes are found to be mutated in 30\% of human cancers \cite{Adjei01,Karnoub08,Cheng08}, 
making their products extremely important therapeutic targets \cite{Bos89}. 
While the intracellular biochemistry of Ras genes is well documented, the biophysical 
mechanism and role of Ras clustering in the plasma membrane remains little understood.

Ras GTPases are small (21 kDa), lipid-anchored peripheral membrane proteins involved 
in signal transduction \cite{Adjei01}.  Three Ras isoforms H-Ras, K-Ras and N-Ras are expressed 
in all mammalian cells. These isoforms contain a conserved G-domain which binds guanine 
nucleotides \cite{Abankwa07}. Ras effectively acts as a molecular switch for 
the signal, with ``on'' (GTP-bound) and ``off'' (GDP-bound) states, the former
promoting an association with and activation of effector proteins. 
Although nearly identical with respect to their catalytic and 
effector-binding properties, H-Ras, N-Ras and K-Ras have very different biological 
roles. This functional distinction is believed to result at least in part from 
the differential membrane compartmentalization of Ras isoforms \cite{Hancock03,Henis08}. 
The different distribution of Ras proteins in cellular membranes dictates unique spatio-temporal 
patterns of activation of effector pathways.
A classical example of a pathway involving Ras is the Ras-Raf-MEK-ERK pathway, 
a mitogen-activated protein kinase (MAPK) cascade involved in cell proliferation, 
differentiation, and apoptosis.  
In this pathway, the epidermal growth factor receptor (EGFR), a receptor tyrosine kinase, 
is stimulated. This leads to recruitment and activation of guanine nucleotide exchange factors (GEFs) 
which, by interacting with the Ras G-domains, promote the exchange of GDP for GTP \cite{Abankwa07} 
and lead to Ras activation. Ras$\cdot$GTP activates protein kinase Raf and initiates the 
phosphorylation cascade, ultimately leading to 
double phosphorylated ERK (ERKpp) which then travels into the nucleus and phosphorylates 
transcription factors \cite{Kolch00}. Among other purposes, such cascades can lead to a 
massive amplification of the original signal \cite{Kolch00}.  

Experimental evidence for the formation of nanoclusters (termed clusters from now on) 
is provided by {\it in vivo} and {\it in vitro} experiments. Fluorescence 
resonance energy transfer (FRET) studies show that activation by EGF leads to 
significant decrease in Ras lateral diffusion, suggesting the existence of Ras$\cdot$GTP 
clusters \cite{Murakoshi04}.  A very similar result was obtained by single-molecule
fluorescence microscopy, where GTP-binding of Ras leads to slowly diffusing
active Ras molecules \cite{Lommerse05}. 
Single particle tracking (SPT) studies of fluorescently labeled Ras have also 
demonstrated transient immobility of 
Ras (lasting less than 1s) with high temporal resolution, interspersed with periods of free
Brownian motion \cite{Hancock05}.  Furthermore, 
spatial statistics of fluorescently labeled Raf have shown that Ras and Raf cluster
together \cite{Tian07}.  
It is therefore believed that active Ras forms signaling platforms 
which recruit and activate Raf. As signaling platforms are Ras-isoform specific, the 
signal diversity observed between H-Ras, K-Ras and N-Ras is in part the result of differential 
clustering properties in these isoforms \cite{Prior03}.

\begin{figure}
\includegraphics[width=8cm,angle=0]{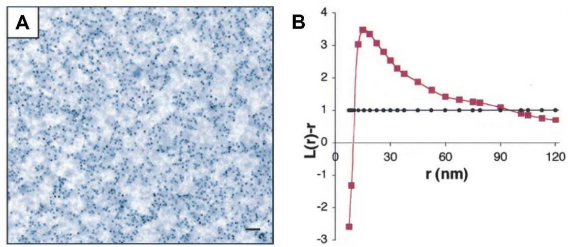}
\caption{Experimental immunoEM data and statistical clustering analysis. 
(A) Electron micrograph of immunogold-labeled Ras domain (GFP-tH where tH is minimal 
plasma membrane targeting motifs of H-Ras) in an \emph{in vitro} plasma 
membrane sheet. Scale bar is 100 nm.
(B) Corresponding point-pattern analysis (red)
and 99\% confidence interval (black).
\copyright Prior {\it et al.} (2003), originally published in 
{\it The Journal of Cell Biology}. doi:10.1083/jcb.200209091 \cite{Prior03}.}
\label{fig:fig1}
\end{figure}

Direct evidence for protein clustering in a membrane can be obtained from high-resolution electron
microscopy (EM). However, Ras is too small and not electron dense
enough to be observed directly.  To circumvent this problem, Prior {\it et al.} used 
GFP-Ras fusion constructs which were treated with gold-labeled anti-GFP antibodies.  
The resulting immunogold point patterns were visualized with EM (immunoEM) 
to quantitatively describe Ras clustering (Fig. \ref{fig:fig1}) \cite{Prior03}.  
It was found that the classical raft model, wherein a fixed number of lipid rafts accommodate a 
fixed fraction of raft-inserted proteins, is incompatible with the observed gold point patterns 
\cite{Plowman05}.  
Indeed, for increasing expression levels, the classical model predicts an increase in the number of 
proteins per raft, and therefore a greater degree of clustering.  
To describe the data, Plowman \emph{et al.} developed an alternative 
raft model, in which the size of Ras clusters remains constant.  Assuming a
constant, expression level-independent ratio of Ras in clusters to Ras monomers 
(cluster/monomer ratio) results in the formation of 
more rafts as expression increases, and supports the notion that lipidated molecules such 
as Ras can drive the formation of rafts in order to create signaling platforms \cite{Plowman05}.  
This alternative model predicts that 40\% of active Ras molecules form clusters 
of radius 6-12nm, each containing about seven Ras molecules, and 60\% are randomly 
distributed monomers \cite{Tian07}.  
While simulations of this model fit immunoEM data, they do not provide 
a biophysical explanation for Ras clustering.  Furthermore, these 
simulations violate laws of equilibrium physics.
Specifically, the law of mass action predicts an increase in the fraction of clustered 
molecules as the expression level is increased (until membrane saturates) \cite{Plowman05}. 
This violation is troublesome as the experiments are done on {\it in vitro} membrane sheets, 
where no active, energy-driven processes can limit cluster size. Membrane sheets were fixed (and
proteins immobilized) after membrane removal from cells \cite{Prior03}, leading to
equilibration of membrane and proteins prior to imaging.

\begin{table}
\centering
\begin{tabular}{|c|c|c|}
\hline
\textbf{Ras density} & \textbf{Ras per} & \textbf{Gold density} \\
\textbf{$\lambda_{\text{ras}}$ ($\mu m^{-2}$)} & \textbf{lattice} & \textbf{$\lambda_{\text{gold}}$ ($\mu m^{-2}$)}\\
\hline \hline
625 & 225 & 264 \\
\hline
1,250 & 450 & 525 \\
\hline
2,500 & 900 & 1,050 \\
\hline
5,000 & 1,800 & 2,100 \\
\hline
\end{tabular}
\caption{Representative Ras densities with corresponding numbers of Ras molecules 
on discretized lattice  membrane as well as gold densities. Shown are the four 
Ras densities used in Figs. \ref{fig:fig4}, \ref{fig:fig5}, and \ref{fig:fig6}. For 
lattice parameters, see {\it Methods}.}
\label{tab:tab1}
\end{table}

Recent experiments even go further and probe the design principles of signaling
by Ras clusters. Such studies suggest that, in the Ras-Raf-MEK-ERK pathway, 
Ras clusters act as an analog-digital-analog converter, where analog continuous 
EGF input is converted into digital, fully active clusters. The number of fully
active Ras clusters, not the activity of individual Ras molecules,
translates into analog ERKpp output \cite{Harding08a,Harding08b}. 
Specifically, these experiments show that Ras mutants with wide-ranging activities 
lead to the same total cellular ERKpp output \cite{Harding05}. This suggests that 
Ras clusters act as digital nanoswitches, which become fully activated even for 
small inputs. Furthermore, the concentrations of active Ras and ERKpp are
directly proportional to EGF input \cite{Tian07}.  Hence, analog inputs produce 
analog outputs, mediated by digital Ras clusters.

\begin{figure}
\includegraphics[width=8cm,angle=0]{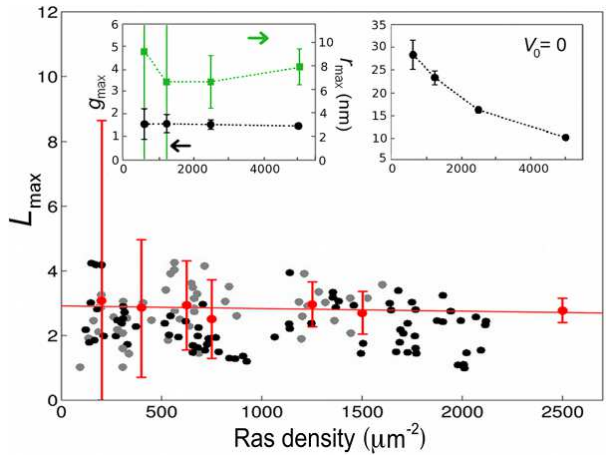}
\caption{Relation between $L_{max}$ and Ras density $\lambda$ for immunoEM data 
of gold labeled GFP-tH (black symbols) and RFP-tH (gray symbols), 
simulation averages and 99\% confidence intervals (red), as well as a linear 
least-squares fit to simulation averages (red line). 
$L_\text{max}$ data points were extracted from Ref. \cite{Plowman05} with \texttt{IMAGE J}.
(Left inset) $g_\text{max}$ (black) and $r_\text{max}$ (green) as a function of $\lambda$.
(Right inset) $L_\text{max}$ as a function of $\lambda$ without long-range repulsion ($V_0=0$).
Error bars represent standard deviations. For simulation details, including
calculation of confidence intervals, see {\it Methods}.
}
\label{fig:fig2}
\end{figure}

Here we consider a physically-motivated model to study Ras clustering. The model 
mainly depends on a close-contact, attractive interaction between active Ras molecules 
(short range $\sim 2$nm) and a repulsive interaction between Ras molecules irrespective
of activity (long range $\sim 5$nm). The short-range 
attraction promotes clustering of active Ras, while the long-range repulsion limits
cluster size. Contrarily to previous models, we make no assumption about a 
constant, expression-level independent cluster/monomer ratio or the existence of 
lipid rafts, thus circumventing controversy surrounding their actuality. We equilibrate
a discretized lattice membrane, occupied with active and inactive Ras molecules, using 
Monte Carlo simulations. After gold-labeling of Ras molecules from simulation 
outputs, we perform a statistical clustering analysis.
The obtained statistical properties of Ras molecules quantitatively agree with 
the statistical properties of immuno-gold point patterns for wide-ranging Ras 
expression levels (Fig. \ref{fig:fig2}) \cite{Plowman05}. 
Our model makes predictions about the signaling properties of Ras clusters, 
supporting the notion that Ras clusters indeed act as an analog-digital-analog converter
\cite{Tian07}.

\begin{figure}[t]
\includegraphics[width=8cm,angle=0]{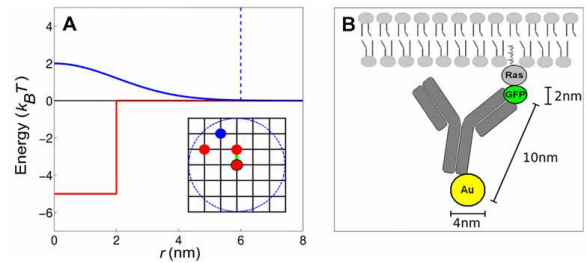}
\caption{Model ingredients. (A) Short-range attraction (red) and long-range repulsion (blue) 
as a function of distance between two Ras molecules for the parameters given in
{\it Methods}. Also shown is the cut-off beyond which the repulsive energy is set to zero 
(blue dashed line). (Inset) 
Representative part of lattice membrane showing three active Ras molecules (red) and one 
inactive Ras molecule (blue). Neighboring active Ras molecules interact via the attractive 
short-range interaction (green bar). 
The cut-off used for the long-range repulsion is representatively shown for the central Ras 
(blue dashed circle).
(B) Schematic of a gold-labeled antibody associated with a GFP-Ras molecule in the 
inner leaflet of the plasma membrane.}
\label{fig:fig3}
\end{figure}

\section{Results}

Prior {\it et al.} \cite{Prior03} studied Ras clustering in plasma membrane sheets using
immunoEM of gold-labeled Ras molecules (Fig. \ref{fig:fig1}A). Gold point
patterns were analyzed based on Ripley's $K$ function. Specifically, the non-linear
transformation $L(r)-r$ was applied where $r$ is the distance between 
gold particles. This function is zero for complete randomness, positive 
for clustering, and negative for depletion (Fig. \ref{fig:fig1}B). 
Plowman {\it et al.} \cite{Plowman05} used the function's maximal value, termed $L_{max}$
for short, as summary statistics for clustering and found that $L_\text{max}$ is independent of 
Ras expression level (Fig. \ref{fig:fig2}, symbols). This was rationalized by an ad hoc clustering model, 
assuming a constant cluster/monomer ratio. Analysis of immuno-gold patterns is consistent with
small clusters, containing approximately 6 to 8 molecules. 
Here we use a biophysical model of Ras clustering in the plasma membrane. 
In our model, a Ras molecule can be in either an active (on) or an inactive (off) state, 
corresponding to the respective GTP-bound and GDP-bound molecules for wild-type Ras.
Both active and inactive Ras are associated with membrane in line with
experimental observation \cite{Philips05}.
The equilibrium probability of a single Ras molecule to be active depends on 
the effective free-energy difference between the on and off states, 
which in turn depends on input signals.
We assume that active Ras molecules experience a short-range attraction, driving 
cluster formation of active Ras, whereas a long-range repulsion limits cluster size 
(Fig. \ref{fig:fig3}A, main panel). Such a long-range interaction may result from 
lipid-anchor induced membrane deformations. 
To obtain equilibrium properties, we approximate the membrane by a square lattice, populated by 
Ras molecules of a specified density (Fig. \ref{fig:fig3}A, inset), and perform 
Monte Carlo simulations. For comparison with immunoEM experiments, 
we added 10nm-long gold-labeled antibodies (maximally one
per Ras) to the Ras molecules in the experimentally observed capture ratio (Fig. \ref{fig:fig3}B). 
We mainly use the four Ras densities given in Table \ref{tab:tab1}.  To 
specifically compare with experiments on varying Ras-expression level (symbols in Fig. \ref{fig:fig2}, 
main panel), we calculate the $L_\text{max}$ value for additional Ras densities. Note that 
these experiments are based on the lipid anchor of H-Ras (tH), which has similar 
clustering properties as active H-Ras \cite{Plowman05}. For details on the 
experiments and our approach, see {\it Methods}.

\begin{figure}
\includegraphics[width=8cm,angle=0]{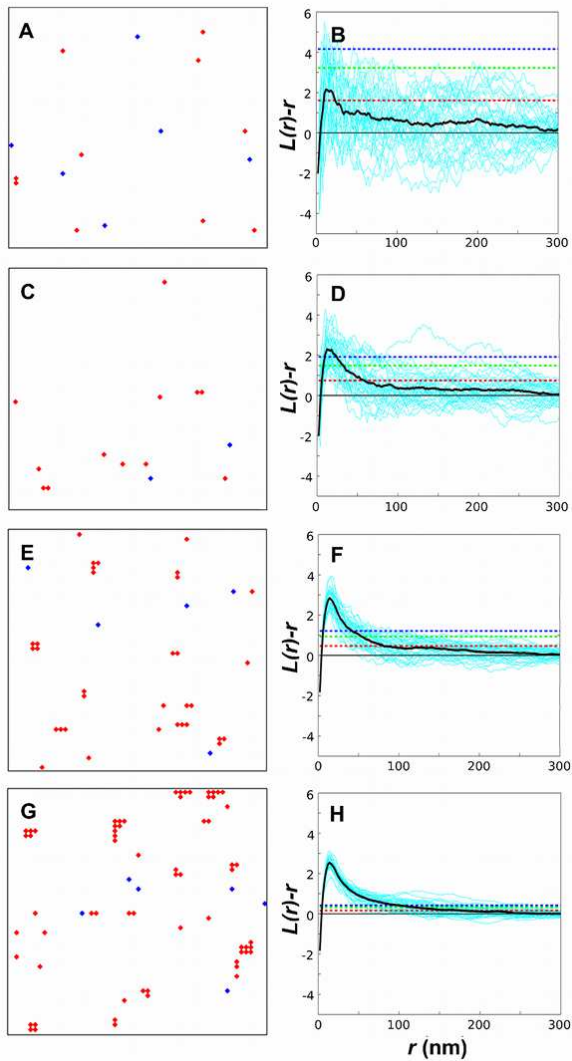}
\caption{Monte Carlo simulations and point-pattern analysis.
Snapshots of equilibrated Ras molecules on lattice membrane (left column; active Ras in 
red and inactive Ras in blue) and corresponding $L(r)-r$ plots (right column) after gold labeling
for the four densities from Table \ref{tab:tab1} (density of Ras molecules increases from top to bottom). 
Shown in the $L(r)-r$ plots are individual simulations (cyan curves),
their averages (thick black curves), as well as 68.3\%, 95.4\%, and 99.0\% confidence intervals 
(red, green, and blue dashed lines, respectively). For simulation details, including
calculation of confidence intervals, see {\it Methods}.}
\label{fig:fig4}
\end{figure}

\begin{figure}
\includegraphics[width=8cm,angle=0]{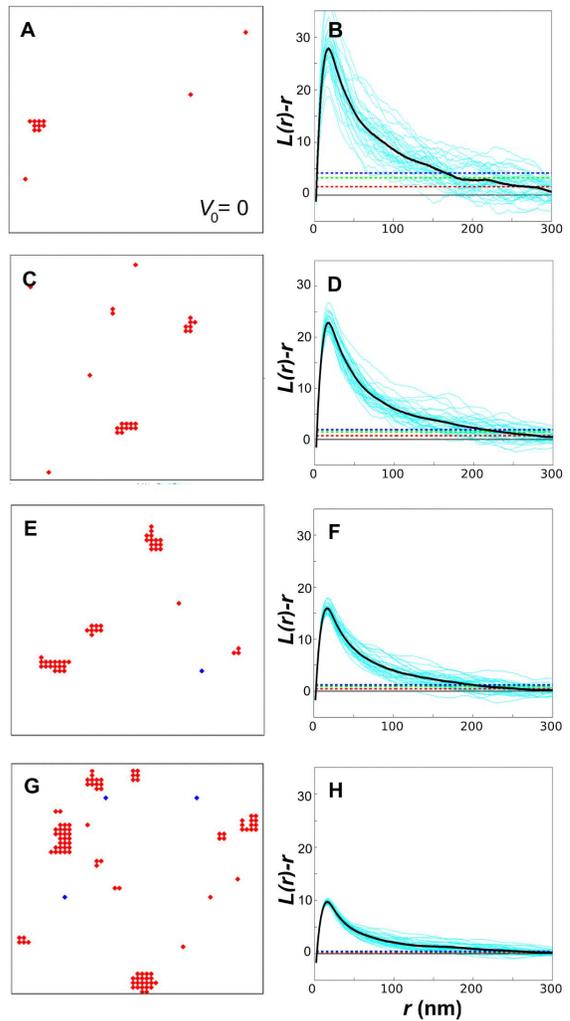}
\caption{Monte Carlo simulations and point-pattern analysis for 
conventional clustering model without long-range repulsion ($V_0=0$). 
For a description of symbols and lines, see Fig. \ref{fig:fig4}. 
}
\label{fig:fig5}
\end{figure}

Figure \ref{fig:fig4} shows typical equilibrated membrane lattices for the four Ras densities 
(left panels) with the corresponding plots of $L(r)-r$ (right panels). For the lowest density, 
individual $L(r)-r$ plots are highly variable. To produce meaningful statements about clustering 
we also show the averaged plot, as well as provide confidence intervals.  In line with 
experiment on varying Ras-expression level, we observe that for our model, 
$L_{max}$ is approximately independent of Ras density (Fig. \ref{fig:fig2}, main panel). 
The same is true if the analysis is done directly on 
Ras molecules instead of the gold particles, demonstrating the robustness of
the result with respect to the details of Ras labeling by gold.   
Distance $r_\text{max}$, defined as the distance corresponding to $L_{max}$, 
is about 8nm (Fig. \ref{fig:fig2}, left inset), or equivalently, 4 Ras molecules. 
Hence, clusters contain few, about 4 to 10, Ras molecules.
An alternative clustering analysis based on the pair-correlation function $g(r)$ 
gives similar results, i.e. $g_{max}$ values are independent of $\lambda$ (Fig. \ref{fig:fig2}, 
left inset). Hence, cluster sizes and their dependence on expression level
are in good agreement with previous estimates \cite{Plowman05}.

We also explored a more conventional clustering model without the long-range 
repulsion, but maintaining the short-range attraction. As shown in 
Fig. \ref{fig:fig5} (left panels), Ras molecules form increasingly larger clusters at 
increasing Ras densities. 
Examination of the $L(r)-r$ plots (right panels) shows that 
$L_{\text{max}}$ decreases for increasing Ras densities (Fig. \ref{fig:fig2}, 
right inset), which is in stark contrast to experiments. Hence, limiting the
cluster size by the long-range repulsion is a necessary ingredient to correctly 
describe immunoEM data and, hence, Ras clustering.

\begin{figure}[t]
\includegraphics[width=8cm,angle=0]{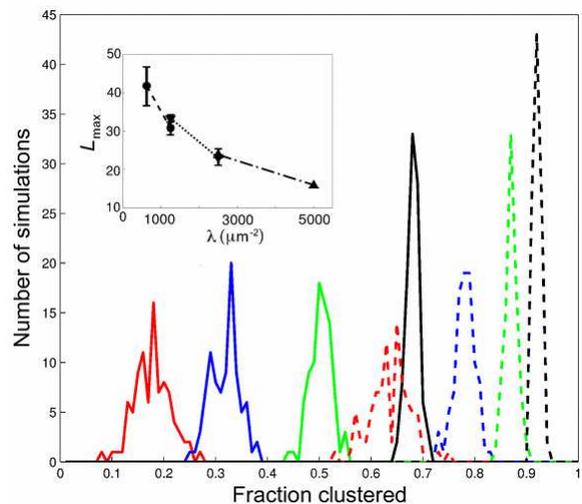}
\caption{Distributions of Ras fractions in clusters.
Different colors correspond to the four Ras densities from Table \ref{tab:tab1}, i.e.
$\lambda=625$ (red), $\lambda=1250$ (blue), $\lambda=2500$ (green), $\lambda=5000$ (black)
in units of $\mu\text{m}^{-2}$.
Shown are results with (solid lines) and without (dashed lines) long-range 
repulsion. A Ras cluster is defined as two or more connected Ras molecules.
(Inset) $L_\text{max}$ for pairwise constant fractions (overlapping fractions), 
i.e. fraction range 0.72-0.75 for $\lambda=625$ and $1250$ (circles and dashed line),
fraction range 0.81-0.84 for $\lambda=1250$ and  $2500$ (triangles up and dotted line), 
and fraction range 0.88-0.91 for $\lambda=2500$ and $5000$ (triangles down and dashed-dotted line).
}
\label{fig:fig6}
\end{figure}

Next we examined the fraction of Ras molecules in clusters. Previous models assumed 
that the fraction is constant, {\it i.e.} independent of Ras density. In contrast, 
Fig. \ref{fig:fig6} shows that in our model the distribution of the fraction clustered increases
significantly with density, indicating that a constant cluster/monomer ratio is not required
to describe the immunoEM data in Fig. \ref{fig:fig2}. Also 
shown in the Fig. \ref{fig:fig6} is the fraction clustered for the conventional clustering model 
without the long-range repulsion. The distribution also shifts to higher values with density, 
although to a lesser extent as fractions are much higher to start out with due to the missing 
long-range range repulsion. To clearly rule out the conventional clustering model
as a suitable model, we tested whether assuming a 
constant fraction clustered (or equivalently, a constant cluster/monomer ratio)
can explain the immunoEM data. For this purpose we collected
simulations from different densities but same fraction clustered and compared their 
$L_\text{max}$ values. However, even with this strong selectivity of simulations, 
$L_\text{max}$ values continued to decrease with increasing density (Fig. \ref{fig:fig6}, 
inset). 

\begin{figure}
\includegraphics[width=8cm,angle=0]{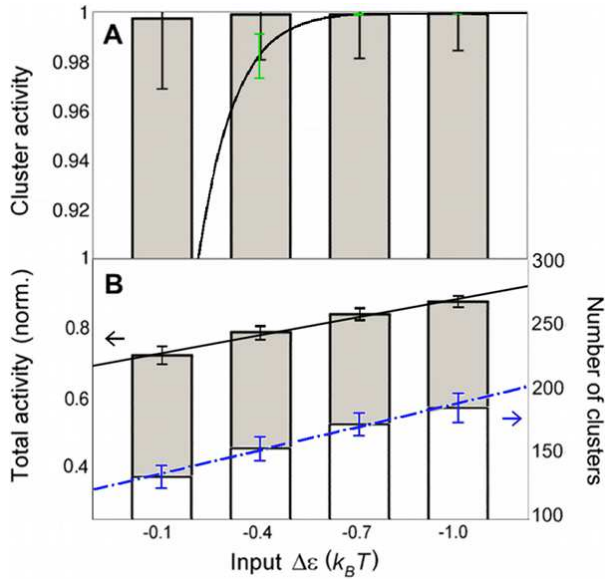}
\caption{Signaling properties of Ras clusters.
(A) Cluster activity as a function of input (parameter $\Delta\epsilon$). 
Cluster activity is defined as fraction of active Ras in clusters from simulations (bar chart),
where a cluster contains two or more contacting Ras molecules.
Also shown is approximate cluster activity $P_\text{on}=1/[1+\exp(N\Delta\epsilon)]$, which assumes that
all $N$ Ras molecules in a cluster (here chose $N=10$) are tightly coupled and hence are 
either all on (active) or all off (inactive) together (black line). 
Black error bars show standard deviation and represent intrinsic noise.
Green error bars represent extrinsic noise, calculated with noise propagation formula
$\delta P_\text{on}=(dP_\text{on}/d\Delta\epsilon)\delta[\Delta\epsilon]=
NP_\text{on}(1-P_\text{on})\delta[\Delta\epsilon]$ for 
$\delta[\Delta\epsilon]=0.05 k_BT$.
(B) Total activity of all Ras molecules in the membrane, normalized by the total number of 
Ras molecules (grey bar chart, left axis), and number of Ras clusters (white bar chart, 
right axis). Also shown are linear fits. Error bars represent standard deviations. To
enlarge black error bars for better visualization in {\it B}, we used the square-root of the total 
variance from pooled simulations of inputs $\Delta\epsilon$, $\Delta\epsilon+0.1 k_BT$, 
and $\Delta\epsilon-0.1 k_BT$.
}
\label{fig:fig7}
\end{figure}

\begin{figure}
\includegraphics[width=8cm,angle=0]{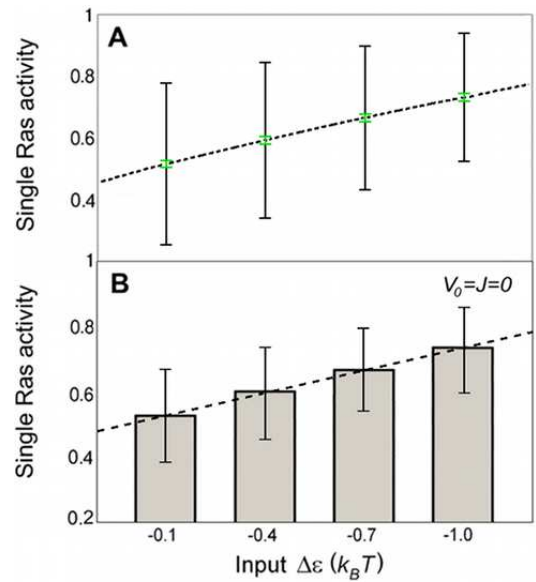}
\caption{Signaling properties of non-interacting Ras molecules.
(A) Activity of single Ras molecule (dashed line; calculated with Eq. \ref{eq:Pon}
for $P_\text{on}$) as a function of input (parameter $\Delta\epsilon$). 
Black error bars represent intrinsic noise, calculated from the square-root of the
binomial variance $P_\text{on}(1-P_\text{on})$.
Green error bars are approximately 0.01 in magnitude and represent extrinsic noise, 
calculated with the noise propagation formula
$\delta P_\text{on}=P_\text{on}(1-P_\text{on})\delta[\Delta\epsilon]$ 
and $\delta[\Delta\epsilon]=0.05 k_BT$.
(B) Total activity of all Ras molecules in the membrane, normalized by the total number of 
Ras molecules (bar chart) and linear fit (dashed line). 
Error bars represent standard deviation, calculated from the square-root of the total variance from 
pooled simulations of inputs $\Delta\epsilon$, $\Delta\epsilon+0.1 k_BT$, 
and $\Delta\epsilon-0.1 k_BT$. 
}
\label{fig:fig8}
\end{figure}

Figure \ref{fig:fig7} shows the signaling characteristics of Ras clusters for 
four different inputs. For input we use the free-energy difference between on (active)
and off (inactive) Ras states ({\it cf.} Eq. \ref{eq:Pon}).
To test if our model produces digital-like nanoswitches, which are fully active even
for small inputs, we identified clusters of two or more connected Ras molecules and calculated
the cluster activity, {\it i.e.} the fraction of active Ras molecules in clusters. 
The bar chart in Fig. \ref{fig:fig7}A shows that our model indeed produces nanoswitches, 
which are fully active even for small stimuli, as indicated by experiments \cite{Harding05}. 
In contrast, the activity of a single Ras molecule does not behave like a switch (Fig. 
\ref{fig:fig8}A, dashed line). Furthermore,
Fig. \ref{fig:fig7}B provides the total activity of all Ras molecules in the membrane 
irrespective of whether Ras molecules belong to clusters or not. We find 
the total activity is approximately proportional to the input (black line) in the range considered
here in line with experiment \cite{Tian07}. In our model, this is due to the fact that 
the number of Ras clusters is proportional to the input (Fig. \ref{fig:fig7}B, blue line).

There has recently been immense interest in understanding the effect of
noise in signal transduction \cite{Rosenfeld05,Bialek08}. 
Biochemical reactions are inherently noisy as they are based on random collisions 
of molecules. This intrinsic noise is further enhanced by the small number of molecules involved. 
Furthermore, rate constants may fluctuate, as they depend on external conditions such as other
molecules not explicitly considered as part of the biochemical reactions. This extrinsic noise 
also includes fluctuations in the input itself. To address how Ras signaling is affected by 
noise, we compare signaling by Ras clusters (Fig. \ref{fig:fig7}) with signaling by non-interacting 
Ras molecules without clustering ability (Fig. \ref{fig:fig8}). Intrinsic noise is inherently part 
of our simulations as Ras is allowed to randomly switch between the active and the inactive states. 
The intrinsic noise for the activity of Ras in clusters (Fig. \ref{fig:fig7}A, black error bars) 
is significantly less than for the activity of a single Ras molecule  (Fig. \ref{fig:fig8}A, 
black error bars) since clusters are fully active and hence suppress random switching. This
difference in intrinsic noise is reduced when considering the intrinsic 
noise of the total activity from all Ras molecules in the membrane, which is only slightly 
smaller for Ras clusters (Fig. \ref{fig:fig7}B, black error bars) than for non-interacting ($V_0=J=0$) 
Ras molecules (Fig. \ref{fig:fig8}B, black error bars). This is due to
the fact that the number of Ras clusters, which is necessarily smaller than the number of Ras 
molecules, can fluctuate significantly (Fig. \ref{fig:fig7}B, blue error bars).
Most importantly, Ras clusters are more robust to input noise, at least for sufficiently large inputs, 
than non-interacting Ras molecules (by comparison of green error bars in Fig. \ref{fig:fig7}A 
and Fig. \ref{fig:fig8}A). Here, input noise represents fluctuations in input much faster than 
assembly/disassembly of clusters but slower than Ras signaling. Therefore, it is assumed that 
extrinsic noise only affects the activity of Ras molecules, not clustering itself 
(see captions of Fig. \ref{fig:fig7}A and Fig. \ref{fig:fig8}A for details).
This shows that Ras clusters have superior signaling properties compared to non-interacting Ras 
molecules without clustering ability.

\section{Discussion}
Different Ras isoforms are known to form nonoverlapping signaling
clusters \cite{Henis08,Eisenberg08}, important for localized signaling of the Ras-Raf-MEK-ERK 
pathway \cite{Sawano02,Reynolds03}, involved in cell proliferation, differentiation, and apoptosis 
\cite{Kolch00}. In addition to the fundamental importance of Ras in this pathway, Ras
mutations are found in 30\% of human cancers \cite{Adjei01,Karnoub08,Cheng08}. 
Ras clusters are also considered evidence of lipid rafts \cite{Plowman05}. Lipid rafts have
attracted considerable interest due to their alleged role in protein sorting and specificity of 
signaling \cite{Pouyssegur02,Marguet06,Morone06}. In this work, we provided a 
biophysical model of Ras clustering, and compared results with gold-point patterns obtained 
from immunoEM  of plasma membrane extracts (Fig. \ref{fig:fig1}). In particular, 
we obtained that clustering of Ras molecules, i.e. the cluster/monomer ratio, 
is independent of expression level (Fig. \ref{fig:fig2}), 
in line with experiments on the lipid anchor of H-Ras (tH) \cite{Plowman05}. 
In our model, as well as in experiments, clustering is quantified by the
maximum value (termed $L_\text{max}$) of function $L(r)-r$ \cite{Plowman05}, where $r$ is the 
distance between gold particles. Our model has two main ingredients exemplified in Fig. \ref{fig:fig3}: 
(1) a short-range attraction between active Ras molecules (e.g. Ras$\cdot$GTP) promoting clustering, and
(2) a long-range repulsion between Ras molecules, which limits cluster size.

Another important feature, which makes our model fundamentally different from previous 
Ras clustering models \cite{Plowman05,Tian07}, is that the fraction of clustered 
Ras molecules is not a model parameter \cite{Plowman05} but a prediction from our 
simulations.  Indeed, if we calculate the fraction of clustered molecules for the 
four densities from Table \ref{tab:tab1}, we 
obtain the distributions shown in Fig. \ref{fig:fig6}. The fraction of
clustered Ras increases with density, indicating that the assumption of
a constant cluster/monomer ratio \cite{Plowman05,Tian07} is misleading
for describing the immunoEM data for different expression levels 
\cite{Plowman05}. Since this assumption violates equilibrium thermodynamics, 
our model is more suitable for describing {\it in vitro} immunoEM data
in absence of energy sources from the cell. Note that in living cells clustering
may in part be regulated by active, energy-dependent mechanisms. For instance,
clustering and signaling of constitutively active K-RasG12V depends on the 
presence of actin fences \cite{Plowman05}. Such membrane-associated actin 
filaments, part of the actin cortex, are highly dynamic and, hence, K-Ras clustering 
may be regulated. An expression level-independent cluster/monomer ratio has also been
found for glycosyl-phosphatidylinositol-anchored proteins (GPI-AP) {\it in vivo}
\cite{Sharma04}. Finally, clustering of proteins in the immunological synapse is 
an active, actin-myosin dependent process \cite{Chichili07}, presumably to overcome the 
entropic barrier of localizing proteins \cite{Weikl04}.

What is the role of Ras clusters beyond simple protein sorting?
Harding {\it et al.} argued that Ras clusters allow for highly precise coding 
of time-dependent inputs, termed high fidelity signaling \cite{Harding08a,Harding08b}. 
First, Ras is highly abundant in the membrane (tens of thousands molecules), 
hence the number of active clusters can be wide-ranging depending on input, 
e.g., EGF. Second, clusters have a short life time (about $0.4$ s). This high
turnover of clusters allows for high temporal precision. Third, Ras 
clusters act as digital nanoswitches, which may lead to noise reduction in the 
signal transmission step across the membrane due to coarse graining and averaging of 
rapidly fluctuating signals. In fact, Ras clusters have similarity to an 
analog-digital-analog converter known from engineering, transmitting 
analog EGF input into analog ERKpp output \cite{Harding08a,Harding08b}.
This design principle was indeed recently confirmed by experiments on the 
Ras-Raf-MEK-ERK pathway in baby hamster kidney cells.
In particular, experiments showed that Ras clusters (actually Raf*-tH with varying kinase 
activity) are fully active even for small inputs (kinase activities) \cite{Harding05}, and 
that the total amount of active Ras (concentration of K-Ras$\cdot$GTP) and ERKpp are 
proportional to EGF input \cite{Tian07}. Such analog ERK activation was recently also 
observed in proliferating mammalian fibroblasts \cite{Mackeigan05}.

The above listed properties of Ras clusters are supported by our model. 
According to Fig. \ref{fig:fig7}A, Ras clusters are fully active, even for small inputs. 
Nevertheless, the number of Ras clusters and hence the total activity of all Ras molecules in 
the membrane are approximately proportional to the input (Fig. \ref{fig:fig7}B), allowing
faithful transmission of continuous, time-dependent input signals. Interestingly, the
activity of a single Ras molecule and of non-interacting Ras molecules are also approximately 
proportional to the input (Fig. \ref{fig:fig8}A and Fig. \ref{fig:fig8}B, respectively). 
However, signaling by Ras clusters is less noisy and, hence, Ras clusters can transmit 
signals more robustly than non-interacting Ras molecules without clustering ability. 
The activity of Ras in clusters 
exhibits smaller intrinsic noise from random switching between active and inactive states 
(Fig. \ref{fig:fig7}A and B, black error bars). The activity of Ras in clusters is also less 
sensitive to extrinsic noise from fast fluctuations in input than non-interacting Ras molecules
(by comparison of green error bars in Fig. \ref{fig:fig7}A and Fig. \ref{fig:fig8}A, respectively). 
The reason for the noise reduction by clusters is that Ras
clusters are fully active, suppressing random switching between active and inactive states, as
well as activity changes due to fluctuations in input.

Our model relies on the short-range attraction and long-range repulsion between
Ras molecules. The physical origin of these interactions are yet to be determined.
However, the attraction may originate from direct Ras-Ras interaction via 
hydrophobic, van der Waals, or electrostatic interactions \cite{Plowman08}, but may also be 
mediated indirectly by scaffold proteins and lipids \cite{Hancock05,Abankwa07,Plowman08}. 
The latter mechanism is supported by the finding that the positively-charged 
polybasic C-terminus of K-Ras binds negatively charged phospholipids and sequesters acidic 
phospholipids, which may attract even more K-Ras molecules. Furthermore, that mutant 
GFP-K-RasG12V S181D has a reduced ability to bind the membrane, as well as 
to cluster \cite{Plowman08}. Such a lipid-mediated mechanism would support the
concept of dynamic lipid rafts, which only form in presence of an activated membrane
protein such as Ras$\cdot$GTP. The physical origin of the repulsion is harder to pinpoint, 
but may result from induced membrane curvature as a result of insertion of the 
farnesyl-polybasic anchor of K-Ras or the farnesyl-palmitate anchor of H-Ras into 
the inner leaflet of the membrane \cite{Abankwa07}. Lipid-anchor induced membrane deformations
are supported by molecular dynamics simulations \cite{Gorfe04}, and may
lead to long-range repulsion \cite{Huang06}. 
Furthermore, recent experiments explicitly
show that smallGTPases (Arf) induce membrane curvature \cite{Lundmark08}.

There are certain shortcomings of the immunoEM data, rendering the cluster analysis 
of immuno-gold data difficult. In our simulations, the addition of gold particles to 
completely random distributions of Ras molecules produced the following. If more than 
one gold-labeled antibody is allowed to bind a Ras molecule, provided there are no 
steric clashes between gold particles, the $L(r)-r$ plot still predicts Ras clustering.  
Varying the antibody length systematically resulted in distance $r_\text{max}$ being
approximately equal to the length of the antibody.  This is due to a small 
fraction of Ras molecules being associated with multiple gold particles: these 
gold particles will be within two antibody lengths of each other, and approximately 
one antibody length from each other on average, resulting in a peak in the 
$L(r)-r$ plot.  This suggests that, unless it is verified that each Ras can 
only bind a single antibody (i.e. the anti-GFP antibody can only bind a single 
epitope on GFP fused to a Ras molecule), immunoEM data can overestimate clustering. 
In contrast, two clustered Ras molecules in contact with each other could each interact 
with separate antigen-binding regions of the same antibody, since an antibody has two 
antigen-binding regions. In this case the cluster of two Ras molecules is unobservable 
by immunoEM, leading to an understimation of clustering in the gold point patterns.  
These issues would have to be addressed if immunoEM studies are to form the basis of 
an accurate quantification of Ras clustering.

In conclusion, a comprehensive description of Ras clustering is an 
essential step in the understanding of Ras signaling properties, and of small, inner-membrane 
GTPases in general. For instance, we showed that clustering leads to robustness to
noise, especially input noise (Figs. \ref{fig:fig7} and \ref{fig:fig8}). While the model we 
have analyzed fits the data (Figs. \ref{fig:fig2} and \ref{fig:fig7}), several questions remain 
unanswered.  For one, lack of high resolution structural information about clustered Ras 
molecules prevents us from ``seeing'' clearly into the physicochemical basis of 
clustering (only partial crystal structures of Ras molecules
exist \cite{Brunger90}). There is also the possibility of 
regulation of clusters from within utilizing specific lipids and scaffold
proteins, which has scarcely been addressed in the literature thus far \cite{Abankwa07}, 
but would provide critical details to the construction of an accurate model 
for clustering.  While we have shown that immunoEM data can aid in the visualization 
of Ras clusters, data regarding the dynamics of clustering are found in the form of 
SPT \cite{Hancock05}, FRET \cite{Murakoshi04}, fluorescence recovery after photobleaching (FRAP) 
\cite{Niv04}, and single-molecule fluorescence microscopy studies \cite{Lommerse05,Lommerse06}.  
These data remain to be integrated into more detailed 
spatio-temporal Monte Carlo simulations, so that the exchange of proteins between
freely diffusing monomers in the membrane and immobile clusters can be investigated
\cite{Nicolau06}. Interestingly, our biophysical model of Ras clustering 
shares the long-range repulsion due to elasticity with recent models of lipid microphase 
separation \cite{Huang06} and chemoreceptor clustering \cite{Endres07,Endres09} in
bacteria. Hence, similar biophysical principles may govern the clustering
of very different types of proteins in prokaryotic and eukaryotic membranes. 
The latter may include EGF and Fc$\gamma$ receptors, which are believed to associate 
with rafts or to form small clusters \cite{Lajoie09,Szymanska09}.

\section{Methods}
\noindent{\bf Experimental immunoEM data}\\
Relevant experiments are described in \cite{Prior03,Plowman05,Harding05,Tian07}. 
Briefly, Ras clustering was examined on intact 2-D sheets of apical plasma 
membrane, ripped off from adherent baby hamster kidney cells directly onto 
EM grids. Ras-fluorescent protein fusion constructs were used, including 
the minimal plasma membrane targeting motif (lipid anchor) of H-Ras fused to 
GFP (GFP-tH) or RFP (RFP-tH), as well as constitutively active H-RasG12V
and K-RasG12V.
These are tagged using affinity-purified polyclonal 
anti-GFP antibodies, conjugated with 4nm gold particles, and
visualized using electron microscopy (immunoEM). The resulting
point patterns of gold particles were analysed for clustering 
(see below for details). Ras isoform clustering was 
found to depend differentially on membrane-associated actin \cite{Plowman05}, 
lipid-raft constituent cholesterol \cite{Prior03}, and scaffold proteins 
galectin-1, galectin-3, and Sur-8 \cite{Parton08}. For a recent review see 
\cite{Henis08}.\\

\ \\ \ \\
\noindent{\bf Biophysical model}\\
A Ras molecule in the membrane can be in either one of two states, 
active (on) with energy $\epsilon_{\text{on}}$ or inactive (off) 
with energy $\epsilon_{\text{off}}$ \cite{Arai06}. 
For wild-type Ras, the active (inactive) state corresponds 
to Ras$\cdot$GTP (Ras$\cdot$GDP). More generally, the two states correspond to 
two different protein conformations, making the two-state model applicable to 
activity mutants and lipid anchors as well.
For any such two level system, 
the probability for a single Ras molecule to be active is 
\begin{equation}
P_\text{on}=\frac{1}{1+e^{\Delta\epsilon}},\label{eq:Pon}
\end{equation}
where $\Delta\epsilon=\epsilon_{\text{on}}-\epsilon_{\text{off}}$ is the free-energy 
difference between the active and inactive states. While 
Eq. \ref{eq:Pon} is not explicitly used for our simulations as it describes the activity of Ras
in absense of interactions, it builds intuition about parameter $\Delta\epsilon$.
This parameter is effectively determined by the input signal of the pathway, 
e.g. EGF, except for the Ras activity mutants and lipid anchors, 
where it desribes an energetic bias in conformational state.

To describe clustering of active Ras molecules, as directly observed 
for K-Ras and H-Ras using {\it in vivo} FRET \cite{Murakoshi04}, 
we introduce short-range attraction $J$ between active Ras molecules, 
driving cluster formation.
In order to limit cluster size, we introduce long-range repulsion $V(r)$, where 
$r$ is the distance between two Ras molecules. 
For the repulsive interaction energy, we use a Gaussian function as previously
applied for describing microphase separation of lipid mixtures \cite{Huang06}
\begin{equation}
V(r) = V_0 \cdot \exp\left(-\frac{r^2}{2\sigma^2}\right), \label{eq:V}
\end{equation}
where $V_0$ is the maximal repulsion for two Ras molecules in close proximity and 
$\sigma$ is the width, i.e. the range of the repulsion beyond which the potential drops 
quickly (Fig. \ref{fig:fig3}A).  The frustration between short-range attraction and 
long-range repulsion leads to small clusters. More precisely, the optimal cluster size 
corresponds to the minimum of the cluster energy divided by the number of Ras molecules 
in the cluster, i.e. the energy density \cite{Ranjan08}. 
The parameters used in this study are $\Delta\epsilon=-0.8$, $J=-5.0$, $V_0=2.0$, and $\sigma = 2$nm.
Long-range repulsions were neglected beyond a 6nm cut-off to reduce the 
calculation time. All energies are in thermal energy units $k_BT$ 
with $k_B$ being the Boltzmann constant and $T$ the absolute temperature.\\

\noindent{\bf Monte Carlo simulations}\\
Since the immunoEM data are obtained from {\it in vitro} membrane sheets, 
clustering is an equilibrium process. Models of such phenomena are therefore 
particularly amenable to Monte Carlo simulations, which include energetics as
well as entropy \cite{Ranjan08}.  To set up simulations, we discretize a 
defined area of the plasma membrane inner leaflet to obtain a two-dimensional $M\times M$ 
square-lattice where $M$ is the lattice size. On the lattice, each position is uniquely described by an 
index $i$ (if the 2-D lattice is thought of as a linear array of length $M^2$). We 
assign a Boolean value $s_i$ to every Ras, where $s_i=0$ if Ras $i$ is active, and $s_i=1$ 
if Ras $i$ is inactive.  Using this notation, we can construct an energy function describing 
the total energy $E$ for a set of $N$ molecules
\begin{equation}
E = \sum_{i=1}^{N}\Delta\epsilon\cdot(1-s_i) + \sum_{\langle i,j \rangle}J\cdot(1-s_i)(1-s_j) + \sum_{i,j}V(r_{ij})
\end{equation}
where $\langle i,j\rangle$ denotes nearest-neighbor pairs. 

After randomly generating the positions of the starting Ras molecules on the lattice, 
individual Ras molecules are chosen at random and attempted to move to a new location 
on the lattice. Included in 
each step is a probability of switching between active and inactive Ras. Moves are accepted or
rejected based on the Metropolis-Hastings algorithm. 
We use a lattice of size $M=300$ and a lattice constant $a=2$nm (the size of a
Ras molecule \cite{Nicolau06}), resulting in a 
$0.36\,\mu$m${}^2$ membrane. In order to reduce boundary effects of the lattice, 
we adopt periodic boundary conditions.\\

\noindent{\bf Gold particles}\\
In order to compare the simulation outputs with immunoEM data, gold-labeled antibodies 
are added to the equilibrated Ras molecules. The length of the antibody used in the experiments
is 10nm \cite{Plowman05}. When an antibody binds a Ras molecule, the 
gold particle associated with the antibody can at any one time occupy any position on 
the surface of a hemisphere around the Ras molecule (Fig. \ref{fig:fig3}B). For 
simplicity, the radius of the hemisphere is chosen equal to the length of the antibody, 
and the centre of the gold particle's position is projected onto the plane of the membrane, 
defining the particle's position on the lattice.  This position is then matched against previous 
gold positions for steric clashes, and if it is found to be closer than 4nm from another gold 
particle, the position is rejected and another Ras is picked at random.  This process 
is iterated until 42\% of Ras molecules are occupied, corresponding to the 
experimentally-observed capture ratio \cite{Plowman05}.\\

\noindent{\bf Cluster analysis}\\
We use three functions to evaluate the degree of Ras clustering: $K(r)$, 
$L(r)-r$ and $g(r)$. Ripley's $K$-function $K(r)$ was first proposed for analyzing 
spatial point patterns \cite{Ripley76}.  $K(r)$ calculates the expected number of 
particles within a distance $r$ of any particle, normalized by the average 
density $\lambda$
\begin{align}
K(r) &= \frac{N(r)}{\lambda} \\
&= \frac{1}{\lambda}\frac{1}{N}\sum_{i=1}^{N}\sum_{i \neq j}^{N}I_{ij}(||x_{j}-x_{i}||\leq r) \\
&= \frac{A}{N(N-1)}\sum_{i=1}^{N}\sum_{i \neq j}^{N}I_{ij}(||x_{j}-x_{i}||\leq r) 
\end{align}
where $A$ is the area of the lattice studied, $N$ the number of Ras molecules or gold 
particles, $\lambda$ the surface density and $I_{ij}(x)$ an indicator 
function which takes a value of 1 if $||x_{j}-x_{i}||\leq r$ and 0 otherwise.  
Under the null hypothesis of complete spatial randomness, $N(r)=\lambda\pi r^2$, 
so $K(r)=\pi r^2$.  An often used non-linear transformation of $K(r)$ 
which we shall employ is  \cite{Prior03}
\begin{equation}
L(r)-r = \sqrt{\frac{K(r)}{\pi}}-r
\end{equation}
which has a value of 0 for complete spatial randomness, is positive for 
clustering, and negative for depletion of particles.  For large $r$, $L(r)-r$ 
is zero on average, since particles are uncorrelated.  
Since $L(r)-r$ is a 
non-linear transformation of $K(r)$, when averaging over multiple simulations
to resemble a large piece of membrane (see below), 
the $K(r)$ values are averaged first and only then transformed into $L(r)-r$. 
For further analyses of simulations, we use summary statistics $L_\text{max}=\max[L(r)-r]$ and
corresponding distance $r_\text{max}=\arg\!\max_r[L(r)-r]$.

The pair-correlation function $g(r)$ can be defined in two different 
ways.  The first by normalizing and differentiating $K(r)$ \cite{Mattfeldt05,Mattfeldt06}
\begin{equation}
g(r) = \frac{1}{2\pi r}\frac{dK(r)}{dr}
\end{equation}
and the second by counting in a similar manner to $K(r)$ but in concentric rings:
\begin{equation}
g(r) = \frac{A}{N(N-1)}\frac{1}{2\pi r a}\sum_{i=1}^{N}\sum_{i \neq j}^{N}I_{ij}(r-a < ||x_{j}-x_{i}||\leq r\big)
\end{equation}
for $r\geq a$, where $a$ is the lattice constant.  Testing these two 
versions of the pair-correlation function yielded slightly different absolute values of 
$g(r)$, but the relative behaviors of the two were identical.  For a random distribution 
of particles and for large $r$ in general, $g(r)$ takes a value of 1 on average. 
We again use summary statistics $g_\text{max}=\max g(r)$ and $r_\text{max}=\arg\!\max_r g(r)$.

To estimate confidence intervals for the $L(r)-r$ cluster analysis, 
99 simulations were run for each density with all 
interactions set to zero, simulating a random 
distribution of Ras in order to obtain an estimate of the background clustering noise 
intrinsic to each density.  Triplets of $K(r)$ were averaged to 
simulate a 1$\mu$m$^2$ membrane as used in experiments, and $L(r)-r$ values were calculated 
for each.  The 68.3\%, 95.4\%, and 99.0\%  confidence intervals for individual measurements 
were obtained by multiplying the standard deviation of the 33 triplets by 1, 2, and 2.576,
respectively.
Standard deviations for $L_\text{max}$, $g_\text{max}$, and $r_\text{max}$ were calculated based 
on triplets as well. \\

\section*{Supporting Information}
\noindent{\bf Accession Numbers}\\
The primary protein accession numbers from the Swiss-Prot databank (http:/\!\!/www.ebi.ac.uk/swissprot/)
for the proteins mentioned in the text are: H-Ras P01112, K-Ras P01116, and N-Ras P01111.\\

\begin{acknowledgments}
We thank Vania Braga, Emmanuelle Caron, and Tony Magee for helpful comments on the manuscript
as well as Suhail Islam for computational support. 
OK and RGE acknowledge funding from the Centre for Integrated Systems Biology at 
Imperial College (CISBIC).  RGE acknowledges funding from the BBSRC grant BB/G000131/1. 
\end{acknowledgments}


\end{document}